\documentclass[pra,twocolumn, superscriptaddress, aps, showpacs, amsmath, amssymb,  nofootinbib]{revtex4}

\usepackage{graphicx}
\usepackage{dcolumn}
\usepackage{bm}
\usepackage{bbm}
\usepackage{amssymb}
\usepackage{amsfonts}
\usepackage{amsthm}
\usepackage{dsfont}

\begin{document}

\title{Codification Volume of an operator algebra and its irreversible growth through thermal processes.}
\date{\today}

\author{Javier M. Mag\'an}\email{javier.magan@iip.ufrn.br}
\affiliation{International Institute of Physics, Universidade Federal do Rio Grande do Norte, 59012-970 Natal, Brazil}

\author{Simone Paganelli}
\affiliation{International Institute of Physics, Universidade Federal do Rio Grande do Norte, 59012-970 Natal, Brazil}


\begin{abstract}
Given a many-body system, we define a quantity, the Codification Volume of an operator algebra, which measures the size of the subfactor of the full Hilbert space with whom a given algebra is correlated. We explicitly calculate it for some limit cases, including vacuum states of local Hamiltonians and random states taken from the Haar ensemble. We argue that this volume should grow irreversibly in a thermalization process, and we illustrate it numerically on a non-integrable quantum spin chain.
\end{abstract}

\maketitle



\section{Introduction}

Thermalization in quantum mechanics has been investigated for a long time. Recently it has increasingly gained attention (\cite{Deutsch1991, page, provepage, srednicki1994} and references therein), and it is now a subject of central interest in different fields of physics, ranging from black holes \cite{hayden,susskind,lashkari2013, us3,stanford2014,esperanza2010, vijay2011} to statistical mechanics \cite{Tasaki1998, Linden2009}, quantum information \cite{Riera2012} and many-body physics \cite{Calabrese2006, Barthel2008, mussardo2009,lea2012, scardicchio2010, brandao2012,vadim,huse,huse2,vadim3}. Generalized Gibbs ensembles \cite{ Gogolin2011,Cassidy2011} have also been studied for systems possessing a greater number of conserved charges. 
 
If the system is in a pure state $\rho=|\psi\rangle\langle \psi| $, a possible way to analyze the thermal
properties of a subsystem $A$ is to study its Entanglement Entropy (EE), defined as 
\begin{equation}\label{eqn:entE}
S=- \textrm{Tr}\rho_A\log\rho_A, 
\end{equation}
where $\rho_A$ is the reduced density matrix of $A$, obtained from $\rho$ by tracing out the degrees of freedom of  the complementary subsystem.

Intuitively, a thermalization mechanism leads to the relaxation of the state of the subsystem to  a stationary thermal density matrix. 
One of the main challenges involves showing how an opportunely defined quantum distance $\textrm{D}(\rho_A(t),\rho_{\textrm{thermal}})$, between the evolving reduced density matrix and 
the associated reduced thermal density matrix, decreases with time.
After some relaxation time $t_{\textrm{r}}$ quantities such as correlations and the EE are expected to exhibit a 
thermal behavior up to corrections related to the size of $\textrm{D}(\rho_A(t),\rho_{\textrm{thermal}})$.
It has been shown \cite{page,provepage} that states with  an \emph{extensive} scaling of the EE not only exist, but they are also typical in the Hilbert space. Thus, the EE appear to be  an appropriate quantity to keep track of a thermalization process.

This approach was considered in \cite{us3} for quantum systems defined on expander graphs \cite{avi1,avi2}. Due to the expansion properties of these type of interaction 
graphs, vacuum \emph{area laws} for EEs are also \emph{volume laws}, with the conclusion that EEs are extensive in the vacuum. Since vacuum states are not 
expected to exhibit thermal behavior, extensivity of EE might be a misleading characterization of thermality for such special situations, as  argued in \cite{us3}.
Notice that the same problem reoccurs in other cases, such as non-commutative quantum field theories \cite{fuertes}.

To better explain the problem, let us consider the simple example of a singlet state in a two spin-$\frac{1}{2}$ system
\begin{equation}\label{twostate}
\vert\psi\rangle=\frac{1}{\sqrt{2}}(\vert\uparrow\downarrow\rangle_{1,2}-\vert\downarrow\uparrow\rangle_{1,2}).
\end{equation}
As it is well known, this is a maximally entangled state and the reduced matrix of one of the two spins is maximally mixed   $\rho_1=\rho_2=\frac{1}{2}\mathds{1}_{2\times 2}$, with EE equal to $ S_{E}= \log 2$. Although the state of the full system is different from the  $2$-spin maximally mixed (or thermal) density matrix, the reduced density matrix is a thermal state at infinite temperature. Now embed state (\ref{twostate}) into a larger Hilbert space of $n$ spins, and consider the state:
\begin{equation}\label{twostateN}
\vert\psi\rangle=\frac{1}{\sqrt{2}}(\vert\uparrow\downarrow\rangle_{1,2}-\vert\downarrow\uparrow\rangle_{1,2}) \otimes \vert\varphi\rangle_{3,\cdots, n}\;,
\end{equation}
where the numbers $1,\cdots, n$ label the spins. The reduced density matrices of the first and second spins are the same as before, but it is clear that information about spin $1$ has not been \emph{thermalized} at all, considering that the spin $1$ is only correlated with the spin $2$. 

In this case, any distance between the reduced density matrix of spin $1$ and its associated thermal density matrix is exactly $0$. Therefore, one needs to ask where the information about a subsystem has been located on the full quantum state in order to conclude that the state is fully thermalized. 
Certainly, in the previous case one could have simply measured the EE of the subsystem formed by spins $1$ and $2$, which is identically zero, to conclude that (\ref{twostateN}) does not produce thermalization for every subsystem. Nonetheless, there are more complex cases in which the EE of every subsystem scales extensively, even if the state is far from being thermal. A deeper analysis is needed. 

The purpose of this article is to define a quantity able to monitor thermalization even in such special situations. We term this quantity the \emph{Codification Volume} of an operator algebra, $\Omega_{\textbf{A}}^{\epsilon}(\rho)$. Due to entanglement in many body quantum physics, information about a given subsystem, which will be defined in terms of its algebra ${\textbf{A}}$, is generically spread over another subsystem $B$, with associated algebra ${\textbf{B}}$. Roughly speaking, $\Omega_{\textbf{A}}^{\epsilon}(\rho)$ will be the dimension of the subsystem $B$ associated to  ${\textbf{B}}$. 

In Sec. \ref{sec:CV} we define $\Omega_{\textbf{A}}^{\epsilon}(\rho)$ and analyze it in different cases, including ground states of local theories and its average over the full Hilbert space. In the first case, when the theory has a gap, we find $\Omega_{\textbf{A}}^{\epsilon}(\rho_{\textrm{vacuum}})\simeq k$, where $k$ is the number of nearest neighbors of each single site. In the second case we find $[\Omega_{\textbf{A}}^{\epsilon}(\rho)]_{\textrm{average}}= \frac{n}{2}$ in the thermodynamic limit, where $n$ is the number of sites of the system. We conclude that this quantity shows a hierarchical jump from thermal to non-thermal states. This demonstrates why vacuum states of quantum systems defined on expander graphs, although satisfying extensivity of EE, are not expected to exhibit thermal behaviour. Information about a single site is stored \emph{locally}, within its nearest neighbors, instead of \emph{globally} within a number of sites scaling as the total number of degrees of freedom.

In Sec. \ref{sec:irrgrow} we argue that $\Omega_{\textbf{A}}^{\epsilon}(\rho)$ grows irreversibly during a thermalization process, from $\mathcal{O}(1)$ to $\mathcal{O}(n)$.  We illustrate these aspects numerically in Sec.  \ref{sec:numerics}, by numerical simulations of a non-integrable quantum spin chain. 

Finally we conclude that the specific functional form of $\Omega_{\textbf{A}}^{\epsilon}(\rho(t))$ in non-equilibrium thermal processes is closely related to the structure of interactions of the theory, reflecting the ability of a system to \emph{hide} information about small subsystems, by correlating it with bigger and bigger subsystems. This provides a different formulation of the so-called ``Fast Scrambling Conjecture'' \cite{hayden,susskind}, parallel to the approach developed in \cite{hayden}. In particular, $\Omega_{\textbf{A}}^{\epsilon}(\rho)$ mathematically formalizes the intuition obtained from thought experiments in black hole physics. In this field, a typical thought experiment consists in `throwing in' a small subsystem $\textbf{A}$, which qualifies as a perturbation of the black hole state.  The combined system is then left to relax  in such a way that any information contained in the small initial subsystem is mixed throughout the large set of internal degrees of freedom of the black hole and radiation \cite{hayden,susskind}, so that this  information becomes only accessible through fine-grained measurements of the final black hole state. We will show how this \emph{information mixing} throughout the whole quantum state is followed by $\Omega_{\textbf{A}}^{\epsilon}(\rho)$.

\section{The Codification Volume of an operator algebra}\label{sec:CV}

In this section, we introduce the concept of Codification Volume, which stands as the core focus of the article.

Consider a many-body system with a Hilbert space having a tensor-product structure of  $n$ single-particle  Hilbert spaces 
\begin{equation}\label{hilbert}
\mathcal{H}=\otimes_{i}\mathcal{H}_{i}~~~~~i=1,..,n \;.
\end{equation}
A $m$-body subsystem  $A$ can be obtained by selecting only $m$ single-particle Hilbert spaces, giving a reduced Hilbert space for $A$ of the form:

\begin{equation}\label{hilbertred}
\mathcal{H}_{A}=\otimes_{l}\mathcal{H}_l~~~~~\left\{l\right\}\subset  \left\{i\right\}\;.
\end{equation}
If the single-particle Hilbert space has dimension $d$, then $\dim \mathcal{H}=d^{n}$ and  $\dim \mathcal{H}_{A}=d^{m}$.

Equivalently, we can define a subsystem in terms of operator algebras. The operator algebra $\textbf{A}_{\mathcal{H}}$ acting on $\mathcal{H}$ can be constructed as:
\begin{equation}\label{algebra}
\textbf{A}_{\mathcal{H}}=\otimes_{i}\textbf{A}_{i}~~~~~i=1,..,n\;,
\end{equation}
where $\textbf{A}_{i}$ is the ``single particle'' operator algebra. In this terms, the operator algebra $\textbf{A}\in \textbf{A}_{\mathcal{H}} $ characterizing the possible measurements performed in subsystem $A$ is of the form
\begin{equation}\label{subalgebra}
\textbf{A}= \mathds{1}\otimes\cdots\otimes \textbf{A}_{l_1}\otimes \mathds{1}\otimes\cdots\otimes \textbf{A}_{l_2} \otimes \mathds{1}\otimes\cdots\otimes \textbf{A}_{l_m} \otimes \mathds{1}\otimes\cdots\otimes \mathds{1}\;.
\end{equation}
If the dimension of the single particle operator algebra is $d$, the dimension of $\textbf{A}$ is $d^{m}$.

Defining a subsystem in terms of corresponding operator algebras is a more general procedure and can also be applied for systems without a Hilbert space of the form~(\ref{hilbert}), a program explored in \cite{amilcar1,amilcar2}. 
In this article, the computations and definitions are restricted to these factorizable cases, but the quantity defined below might be extended to other type of operator algebras by using the framework of \cite{amilcar1,amilcar2}. 

We are interested in the question of whether the information associated to $\textbf{A}$, defining a subsystem $A$, can be ``localized'' in a different $\textbf{B}$, defining a subsystem $B$. 
Even if the subsystem $A$ is entangled with the rest, one might expect that in many cases most of its information is distributed only in a certain subsystem $B$, associated to an algebra $\textbf{B}$. So, fixing an accuracy $\epsilon$,  one can ask if it is possible to find a minimum size $\textbf{B}$ where all the information of $\textbf{A}$ is localized on.
In order to formulate this question quantitatively, we introduce hereafter a quantity $\Omega_{\textbf{A}}^{\epsilon}(\rho)$, called the \emph{Codification Volume} of an operator algebra $\textbf{A}$.

\begin{figure}
  \begin{center}
\includegraphics[width= 5.5cm,angle=0]{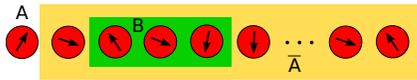}\\
    \caption{Pictorial scheme of the systems considered.
    \label{fig:scheme}}
  \end{center}
\end{figure}

Consider two operator algebras $\textbf{A}$ and $\textbf{B}$, defining two subsystems $A$ and $B$. 
In this article, we will consider $\textbf{A}$ and $\textbf{B}$ with associated disjoint subsystems of a factorizable Hilbert space, where disjoint means that $A$ and $B$ are composed by different single particle factors of~(\ref{hilbert}), as pictorially shown in Fig. \ref{fig:scheme}.
We want to measure the total correlation between the two of them. This translates into finding  the distance of the associated reduced density matrix $\rho_{\textbf{A}\textbf{B}}$ to a factorized state $\rho_{\textbf{A}}\otimes 
\rho_{\textbf{B}}$ \footnote{We want to emphasize that the reduced density matrix of an operator algebra is perfectly defined, and indeed it works nicely for cases in which the operator algebra does not define a partition of the Hilbert space \cite{amilcar1,amilcar2}.}. For this task we use the quantum relative entropy $S(\rho\Vert \sigma)$, defined by
\begin{equation}\label{relativedefinition}
S(\rho\Vert \sigma)=\textrm{Tr}\rho (\log\rho-\log\sigma)\;,
\end{equation}
because it is a higher bound of all other possibilities \cite{eisert,cirac}. In our case, in which $A$ and $B$ are disjoint the relative entropy coincides with the Mutual Information (MI) $I(\textbf{A},\textbf{B})$
\begin{equation}\label{relativeI}
S(\rho_{\textbf{A}\textbf{B}}\Vert \rho_{\textbf{A}}\otimes 
\rho_{\textbf{B}})= S(\rho_{\textbf{A}})+S(\rho_{\textbf{B}})-S(\rho_{\textbf{A}\textbf{B}})= I(\textbf{A},\textbf{B}).
\end{equation}
The MI gives the total amount of correlations between two systems \cite{groisman2005} and it is a measure of how much we can learn about $\textbf{A}$ by studying $\textbf{B}$ and viceversa.
 
Now we define the Codification Volume of an operator algebra $\Omega_{\textbf{A}}^{\epsilon}(\rho)$, for a given accuracy $\epsilon$, as
\begin{equation}\label{omega}
\Omega_{\textbf{A}}^{\epsilon}(\rho)\equiv \log [\textrm{Min}(\textrm{dim} \, \mathcal{H}_{B})]\;,
\end{equation}
where the minimum over all algebras $\textbf{B}$ or, equivalently over all subsystems $B$ with associated dimensionality $\textrm{dim} \, \mathcal{H}_{B}$, is taken such that the following relation holds:
\begin{equation}\label{eq:cond}
I(\textbf{A},\bar{\textbf{A}})-I(\textbf{A},\textbf{B})\leqslant \epsilon \;,
\end{equation}
where $\bar{\textbf{A}}$ is the algebra complementary to $\textbf{A}$.
When condition~(\ref{eq:cond}) holds, we  say that, up to $\epsilon$, all the information shared by $\textbf{A}$ might be found by looking only at $\textbf{B}$.

\subsection{Examples: local theories and random states}\label{sec:ex}

To better illustrate the meaning of the Codification Volume, we will analyze it for some simple cases. 
The easiest example is the factorized state in the local basis. For operators algebras of the type~(\ref{subalgebra}) one obtains trivially $\Omega_{\textbf{A}}^{\epsilon}(\rho)=0$. To recover the information about $\textbf{A}$ we do not need to study any other $\textbf{B}$.

A second trivial example is given by the state (\ref{twostateN}). For the algebra of the first spin we have $\Omega_{\textbf{1}}^{\epsilon}(\rho_{\psi})=1$, where $\rho_{\psi}=\vert\psi\rangle\langle\psi\vert$ and $\vert\psi\rangle$ given by~(\ref{twostateN}). In other words, the MI between spin $\textbf{1}$ and other spins is already maximized when considering spin $\textbf{2}$, i.e $I(\textbf{1},\textbf{2})=I_{\textrm{max}}=2S(\textbf{1})=2\log 2$, and adding more spins do not increase the MI, $I(\textbf{1},\textbf{234}\cdots)=I(\textbf{1},\textbf{2})=I(\textbf{1},\bar{\textbf{1}})$.

As a third case, let us consider local discrete theories with finite correlation length $\xi$. The vacuum states of these models can be efficiently approximated by projected entangled pairs (PEPS), so that each degree of freedom is just entangled with its nearest neighbors \cite{cirac}. Considering $\textbf{A}$ as the algebra of one single site, we expect
$\Omega_{\textbf{A}}^{\epsilon}(\textrm{vacuum})=k$, $k$ being the number of nearest neighbors.
More generically, we expect $\Omega_{\textbf{A}}^{\epsilon}(\textrm{vacuum})=\mathcal{O}(1)$ in the thermodynamic limit to be a characteristic feature of local theories. Notice that quantum systems defined on expander graphs seem no different in this regard to other local theories. We will continue to study these cases in future works.

Notice that $\Omega_{\textbf{A}}^{\epsilon}(\textrm{vacuum})$ might be understood as a definition of a \emph{correlation length} in quantum information terms. Therefore, it would be interesting to compute $\Omega_{\textbf{A}}^{\epsilon}(\textrm{vacuum})$ for critical theories, using the framework of \cite{sodano2011}.

The last case we will consider is a random state over the Haar ensemble. We intend to compute the average value of $\Omega_{\textbf{A}}^{\epsilon}(\rho)$ over the full Hilbert space, with all states $\rho$ having equal weight. 
Let us consider the  Hilbert space $\mathcal{H}$, given by (\ref{hilbert})  with  $n$  two-dimensional systems (qubits), i.e with $\textrm{dim}\,\mathcal{H}_{i}=2$ and $\textrm{dim}\,\mathcal{H}=2^n$ . The average EE for a subsystem $A$ composed of $a$ qubits, with a reduced Hilbert space  $\mathcal{H}_{A}$ and a dimension $\textrm {dim}\,{\cal H}_{A} \, = \, 2^{a} \leq 2^{n/2}$, associated to an algebra $\textbf{A}$, is given by  \cite{page,provepage} 
\begin{equation}\label{page}
S_{a,n}\,=\,\sum\limits_{i=i_0}^{2^{n}}\frac{1}{i}-\frac{2^{a}-1}{2^{n-a+1}},
\end{equation}
with $i_0=2^{n-a}+1$.
Notice that here, contrary  to the original reference \cite{page}, we prefer to work with the number of qubits $a$ and $n$, instead of the dimensions of the Hilbert spaces.  The sum in~(\ref{page}) can be written as:
\begin{equation}\label{harmonicsum}
 \sum\limits_{i=i_0}^{2^{n}}\frac{1}{i}=\sum\limits_{i=1}^{2^{n}}\frac{1}{i}-\sum\limits_{i=1}^{2^{n-a}}\frac{1}{i}=H_{2^{n}}-H_{2^{n-a}}\;,
\end{equation}
where $H_{p}\simeq \log p + \gamma +\frac{1}{2p}$ is the so-called Harmonic number, and $\gamma\simeq 0.577$ is the Euler-Mascheroni constant.

\begin{figure}
 \begin{center}
 \includegraphics[width= 5.5cm,angle=-90]{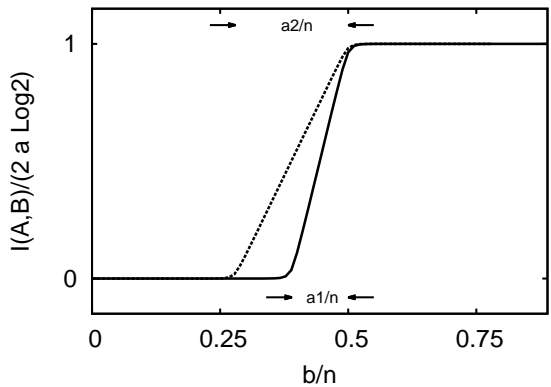}
 \end{center}
    \caption{\small The average Mutual Information between different subsets of the Hilbert space is exponentially suppressed until $b+2a=\frac{n}{2}$. From that point it grows linearly until saturating to its maximum value, given by $I(\textbf{A},\textbf{B})=2a$. The continuos line corresponds to $a_1/n=1/9$, while the dotted line corresponds to $a_2/n=2/9$.}\label{mutualfigure}
\end{figure}

As a result, using formula~(\ref{page}), we can compute the average MI over the Hilbert space. Considering $\textbf{B}$ with $ \textrm {dim}\,{\cal H}_{\textbf{B}} \, = \, 2^{b}$, the structure of the average $I(\textbf{A},\textbf{B})$ over $\mathcal{H}$ is summarized in Fig (\ref{mutualfigure}).

Specifically, for $b\,<\frac{n}{2}$ and  $a+b\,<\frac{n}{2}$ we have:
\begin{equation}\label{caso1}
I(\textbf{A},\textbf{B})\simeq \frac{2^{a+b}-2^{a-b}}{2^{n-a-b+1}}\;.
\end{equation}
 It is interesting to note that when $b=\frac{n}{2}-a-c$, we have $I(a,\frac{n}{2}-a-c)\simeq 2^{-2c}(1-2^{-2a})$. This observation, together with the known bounds relating quantum relative entropy (the MI in this case) and the trace-1 norm \cite{eisert,cirac}, is conceptually similar to the result described in Ref \cite{hayden}, which deals with the factorization properties of quantum states after random unitary evolution.
 
In the region with $b\,<\frac{n}{2}$ and  $a+b\,>\frac{n}{2}$, one has
\begin{equation}\label{caso2}
I(\textbf{A},\textbf{B})\simeq(2(a+b)-n) \log 2\,-\,\frac{2^{3b+a}-2^{2n-a-b}}{2^{n+a+b+1}}\;.
\end{equation}
We conclude that in both previous regimes the MI  is smaller than $I( \textbf{A},\bar{\textbf{A}} )=2 a \log 2$ and keeps growing substantially as we increase $b$.
Finally, when $b\,>\frac{n}{2}$ we obtain
\begin{equation}\label{caso3}
I(\textbf{A},\textbf{B})\simeq  2a \log 2 + \frac{2^{2n-a-b}-2^{2n+a-b}}{2^{n+a+b+1}}\;.
\end{equation}
To apply the definition~(\ref{eq:cond}), we compute the difference between $I(\textbf{A},\bar{\textbf{A}})$ (formula (\ref{caso3}) with $b=n-a$) and $I(\textbf{A},\textbf{B})$
\begin{equation}\label{difference}
I(\textbf{A},\bar{\textbf{A}})-I(\textbf{A},\textbf{B})\simeq g(a)(2^{-2b+n}-2^{-n+a})
\end{equation}
where $g(a)=\frac{2^{a}-2^{-a}}{2^{a+1}}\sim \mathcal{O}(1)$. Setting (\ref{difference}) equal to a finite $\epsilon$,  we obtain the value of
$b=b_{\epsilon}$ corresponding  the CV
\begin{equation}\label{average}
b_{\epsilon}=[\Omega_{\textbf{A}}^{\epsilon}(\rho)]_{\textrm{average}}\xrightarrow{n\rightarrow\infty} \frac{\log(\frac{g(a)}{\epsilon})}{2\log 2}+\frac{n}{2}\simeq \frac{n}{2}.
\end{equation}

\section{The irreversible growth of the Codification Volume during thermalization: Theoretical expectations } \label{sec:irrgrow}

Consider a quantum spin-$\frac{1}{2}$ system in an initial state $|\psi_{\textrm{in}}\rangle$ susceptible of thermalization \footnote{It is clear that not all the states in the Hilbert space will thermalize in physical systems, such as the cases of vacuum states and low lying excitations. This was studied for example in \cite{strongthermal}, where several initial states were found to evolve into different density matrices.}. We expect that after some time $t_{\textrm{scrambling}}$, the evolved state $|\psi (t_{\textrm{scrambling}})\rangle=U(t)|\psi_{\textrm{in}}\rangle$ will have properties similar to those described for the average over the Hilbert space. This characteristic time scale was termed \emph{the scrambling time} in \cite{susskind}. Using the results gained in the previous section we then expect an irreversible growth of $\Omega_{\textbf{A}}^{\epsilon}$, from $\mathcal{O}(1)$ at times $t\sim\mathcal{O}(1)$ to $\mathcal{O}(n)$ at times $t_{\textrm{scrambling}}$. The characteristic time scale for the stabilization of the growth provides an operative definition of $t_{\textrm{scrambling}}$. From this perspective,  a thermal process progresses by codifying any initially localized information in bigger and bigger subsystems.

Here we will consider  the subsystem  $\textbf{A}$  that corresponds to the single spin on site $1$ and study the evolution of $\Omega_{\textbf{A}}^{\epsilon}(\rho(t))$ when starting from two different initial states. The first initial state is:
\begin{equation}\label{af}
|\psi_{\textrm{af}}\rangle=|\uparrow\rangle_{1}\otimes|\downarrow\rangle_{2}\otimes|\uparrow\rangle\otimes\cdot\cdot\cdot|\downarrow\rangle_{n}\;.
\end{equation}
The second initial state is a factorized product of all spins aligned in the positive $y$ direction, called $\vert Y_{+}\rangle$ in \cite{strongthermal}:

\begin{equation}\label{ymas}
|\psi_{Y^{+}}\rangle=\prod_{j=1}^{n}(\frac{1}{\sqrt{2}}(\vert\downarrow\rangle_{j}+i\vert\uparrow\rangle_{j}))\;.
\end{equation}
In the next section, we use a Hamiltonian which will drive the previous initial states to a stationary regime with a structure of MI akin to that of the random state, see Fig (\ref{mutualfigure}) and Fig (\ref{fig:therm}).

The specific form of $\Omega_{\textbf{1}}^{\epsilon}(|\psi(t)\rangle)$ as a function of $t$ and $n$ is directly related to  the interaction structure of the theory. In particular, Lieb-Robinson's causality bounds \cite{lieb} are expected to limit the allowed growth. Let us consider a Hilbert space of $n$ quantum spins $\mathcal{H}=\otimes_{j=1}^{n} \mathcal{H}_j$. To find $\Omega_{\textbf{1}}^{\epsilon}(|\psi(t)\rangle)$, we will compute
\begin{eqnarray}\label{infoarray}
I(\textbf{1},\textbf{2})(t)\nonumber\\
I(\textbf{1},\textbf{23})(t)\nonumber\\
\vdots \nonumber\\
I(\textbf{1},\textbf{23}\cdots\textbf{n})(t),
\end{eqnarray}
where $I(\textbf{1},\textbf{23}\cdots\textbf{k})(t)$ is the MI between the subsystem associated with spin $1$ and the subsystem associated with spins $ij\cdot\cdot\cdot k$. Due to strong subadditivity, notice that:
\begin{equation}\label{infobound}
I(\textbf{1},\textbf{23}\cdots\textbf{n})(t)=I(\textbf{1},\bar{\textbf{1}})=2 S_{\textbf{1}}(t)\geq I(\textbf{1},\textbf{ij}\cdots \textbf{k})(t).
\end{equation}
In addition, Lieb-Robinson bounds force information to propagate inside an effective light cone, up to exponentially suppressed corrections. For a sufficiently short amount of time, the information associated to the first spin is expected to have reached only the second spin. Therefore, considering more spins should not increase the MI, a statement which is expressed by the following equation:
\begin{equation}
I(\textbf{1},\textbf{2})(t)\simeq I(\textbf{1},\textbf{23})(t)\simeq\cdot\cdot\cdot\simeq I(\textbf{1},\bar{\textbf{1}})(t)=2 S_{1}(t)
\end{equation}
At some $t_{1,2}$ the information passes to the third spin. This moment $t_{1,2}$ is signalled by the decrease of $I(\textbf{1},\textbf{2})(t)$. By now the information should be found inside $I(\textbf{1},\textbf{23})(t)$, and adding more spins should not increase the MI, again due to Lieb-Robinson bounds. This in turn means that $\Omega_{\textbf{1}}^{\epsilon}(|\psi(t_{1,2})\rangle)=2$.  The next characteristic time scale is $t_{1,23}$, in which $I(\textbf{1},\textbf{23})(t)$ starts to decrease. By this time $\Omega_{\textbf{1}}^{\epsilon}(|\psi(t_{1,23})\rangle)=3$. This process continues until saturation, in which $\Omega_{\textbf{1}}^{\epsilon}(|\psi(t_{\textrm{scrambling}})\rangle)\sim\mathcal{O}(n)$ . The set of characteristic time scales $t_{1,ij\cdot\cdot\cdot k}$ is clearly related to the ability of the evolution to \emph{hide} the information in bigger and bigger subsystems, by means of the structure of interactions.

In the next section, we give a numerical example to validate this expected behavior.

\section{Numerical study of the growth of the Codification Volume}\label{sec:numerics}

In this section we numerically study the unitary evolution driven by the following Hamiltonian:
\begin{equation}\label{Hamil}
H=-J \sum_{i=1}^{n-1} \sigma_{i}^{z}\sigma_{i+1}^{z}+\frac{J}{2}\sum_{i=1}^{n}  \left(3 \sigma_{i}^{x}-\sigma_{i}^{z}\right ).
\end{equation}
The Hamiltonian couplings are chosen so as to be far from integrability regions \cite{strongthermal}, and the number of spins considered is $n=10$. We study the evolution of the two different states~(\ref{af}) and ~(\ref{ymas}). 
Despite the reduced size of the system, we already observe a relaxation process and the thermalization effects described in the previous sections.

\begin{figure}[h]
\includegraphics[width= 5.5cm,angle=-90]{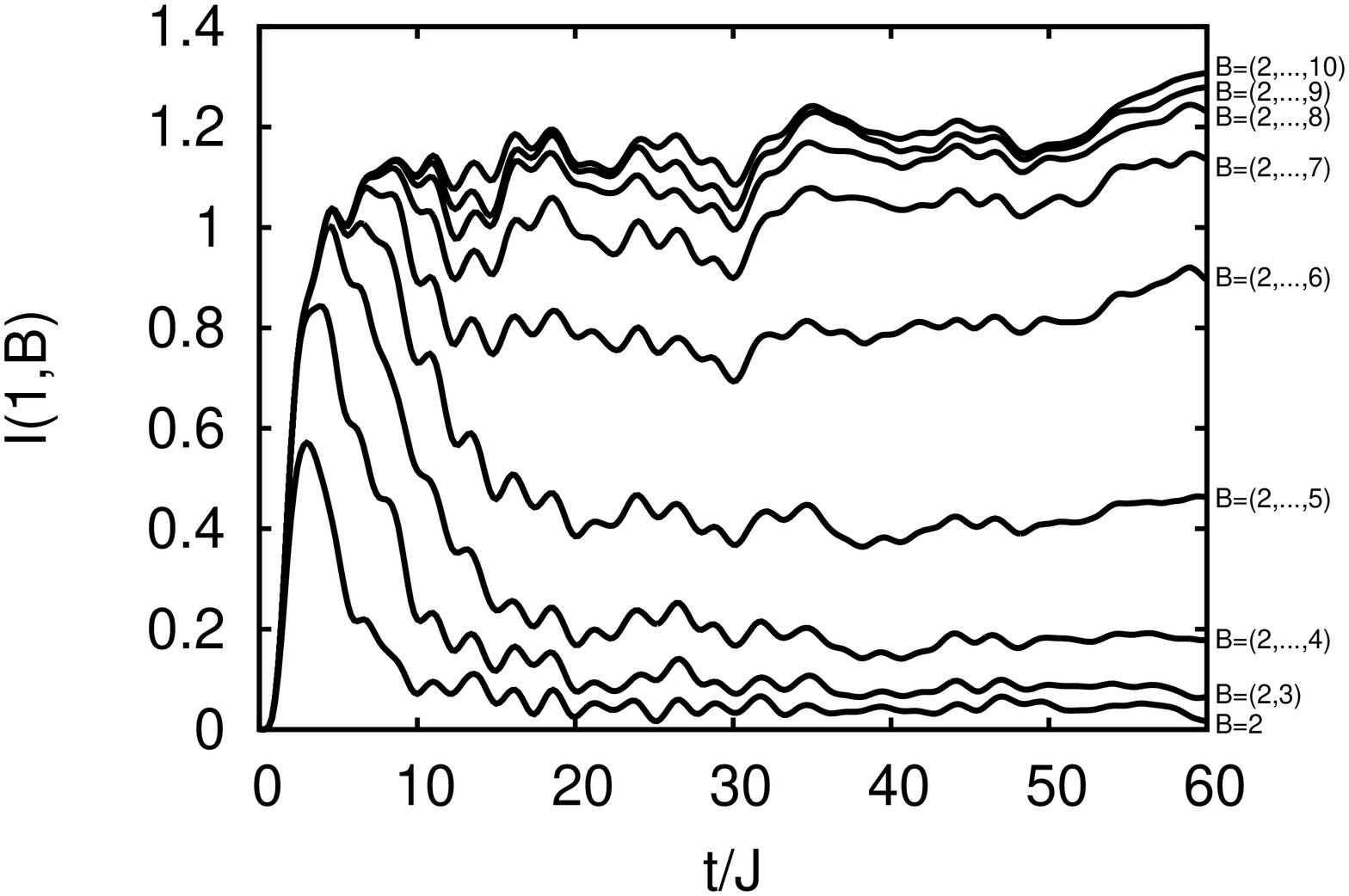}\\
\caption{Temporal evolution of the Mutual Information  between the site $1$ and other possible subsystems of the spin chain. The initial state  is $\vert\psi_{\textrm{af}}\rangle$. }
\label{fig:MIAF}
\end{figure}

\begin{figure}[h]
  \includegraphics[width= 5.5cm,angle=-90]{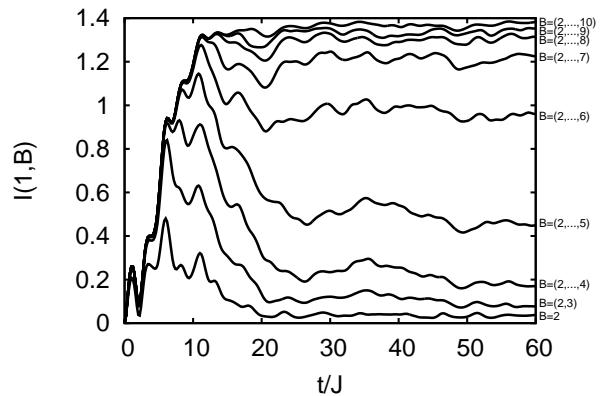}\\
    \caption{Temporal evolution of the Mutual Information  between the site $1$ and other possible subsystems of the spin chain. The initial state  is $\vert \psi_{Y^{+}}\rangle$. }
    \label{fig:MITRAN}
\end{figure}

The evolution of the  MIs between the first site and different blocks of spins is shown in Figs. \ref{fig:MIAF} and~\ref{fig:MITRAN}. The structure reflects the theoretical considerations developed in the previous section. Not only does the information propagate to the right of the chain, but it also gets codified in bigger and bigger subsystems. This can be checked by computing $I(\textbf{1},\textbf{10})(t)$, which approaches monotonically its static value close to zero. Notice that there are slight changes in the relaxation velocity when considering different initial states.

\begin{figure}[h]
\includegraphics[width= 5.5cm,angle=-90]{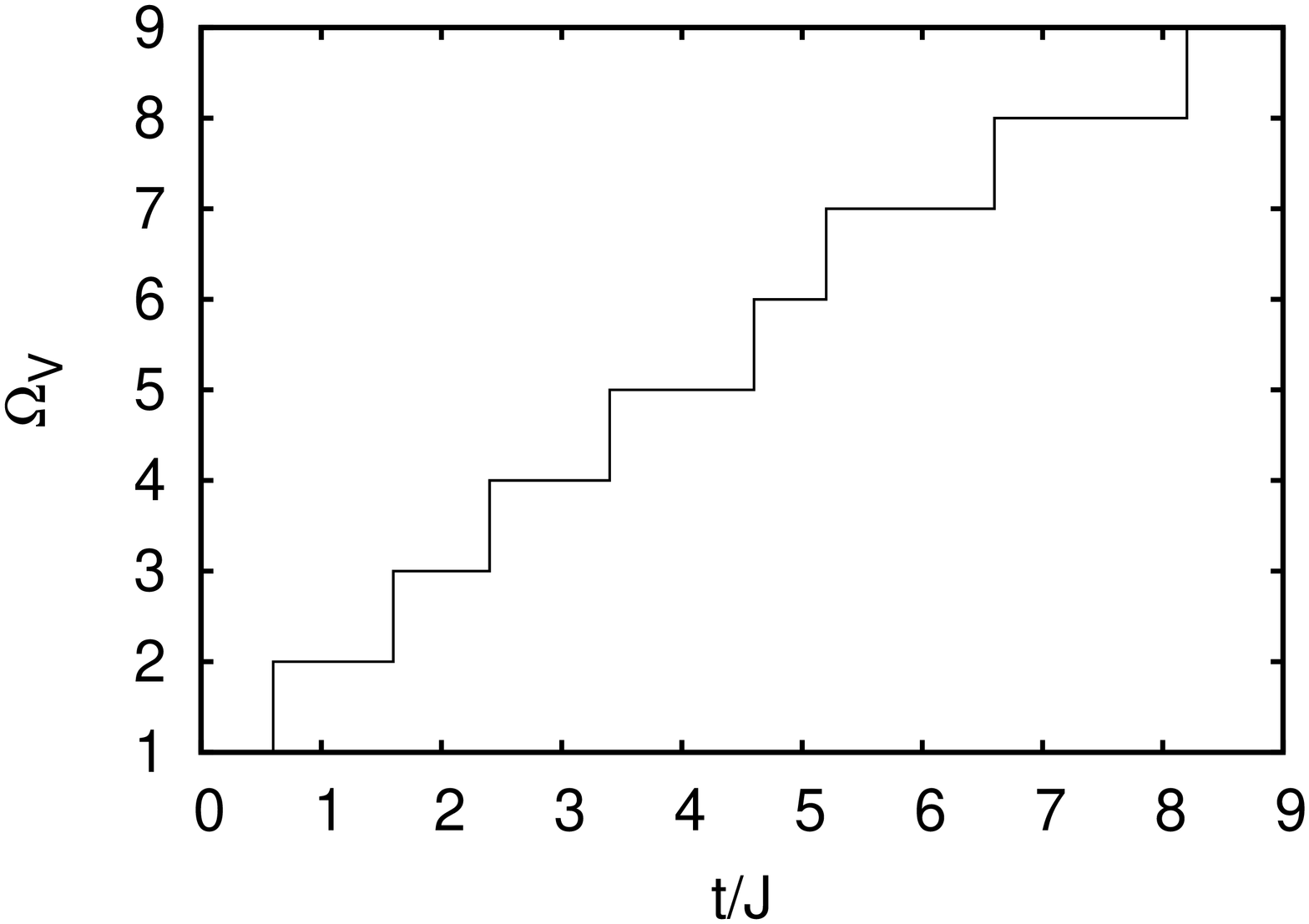}\\
\caption{Temporal evolution of $\Omega_{\textbf{1}}^{\epsilon}(\rho)$,  for $\epsilon=0.0001$ and initial state   $\vert\psi_{\textrm{af}}\rangle$. }
\label{fig:CVAF}
\end{figure}

\begin{figure}[h]
 \includegraphics[width= 5.5cm,angle=-90]{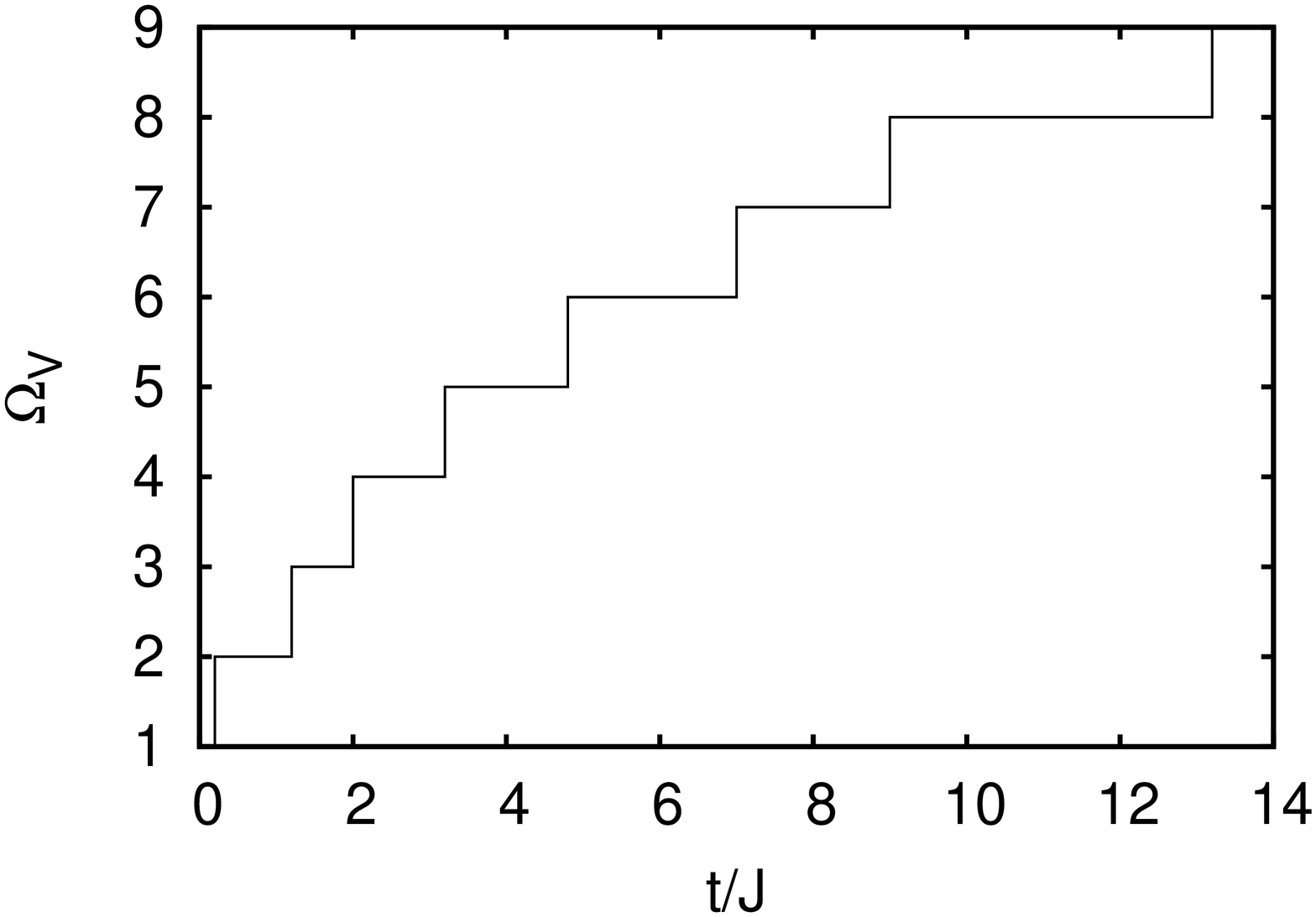}\\
    \caption{Temporal evolution of $\Omega_{\textbf{1}}^{\epsilon}(\rho)$,  for $\epsilon=0.0001$ and initial state  $\vert \psi_{Y^{+}}\rangle$. \label{fig:CVTRAN}}
\end{figure}

The growth of $\Omega_{\textbf{1}}^{\epsilon}(\rho)$ as a function of time is depicted in Figs. \ref{fig:CVAF} and ~\ref{fig:CVTRAN}, for a precision $\epsilon=0.0001$.
The linear growth suggests ballistic propagation of information in the system. Although diffusion is expected for energy equilibration, information stored in entanglement correlations might travel faster through the system, possibly due to decoherence type effects. However, the number of spins is too small to draw such conclusions, and an analysis with larger chains may show a different functional form.

\begin{figure}
  \begin{center}
\includegraphics[width= 5.5cm,angle=-90]{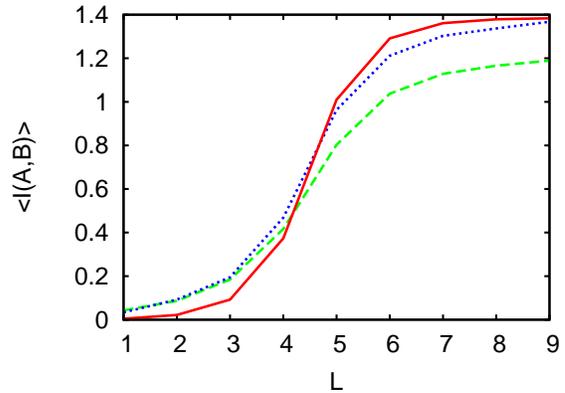}\\
    \caption{(Color online) Average values of the MI for long times.  Initial state $\vert\psi_{\textrm{af}}\rangle$ (green-dashed line) and initial state $\vert \psi_{Y^{+}}\rangle$ (blue-dotted line) are compared with
    a random state (red-continuous line),  \label{fig:therm}}
  \end{center}
\end{figure}

The average values of the MI at equilibrium, together with the theoretical expectation found in Sec.~(\ref{sec:ex}) are plotted in Fig.  \ref{fig:therm}. The structure is the same in the three cases and the quantitative discrepancy is expected. For the theoretical case we are averaging over the full Hilbert space. For the numerical cases energy conservation restricts the average over the full Hilbert space to a \emph{microcanonical} average. It would be interesting to develop an analogue of Page's formula~(\ref{page}) for cases with this type of constrains. Nevertheless, it is interesting to note that the structure at stationarity is very similar for the two non-equilibrium processes, although the intersection of their respective ensembles of states is $\varnothing$, since the initial expectation value of the Hamiltonian is different for each state.

\section{Conclusions}

We have defined a new quantity entitled the Codification Volume of an operator algebra $\Omega_{\textbf{A}}^{\epsilon}(\rho)$. This quantity measures the size of the smallest subsystem needed to recover a specified amount of information about $\textbf{A}$.

By computing the average of $\Omega_{\textbf{A}}^{\epsilon}(\rho)$ over the Hilbert space, we discussed the use of this quantity in studying thermal processes. To compute this average of $\Omega_{\textbf{A}}^{\epsilon}(\rho)$, we used results from \cite{page,provepage}, where the average entanglement entropy of a given subsystem was found. The result is $[\Omega_{\textbf{A}}^{\epsilon}(\rho)]_{\textrm{average}}= \frac{n}{2}$ in the thermodynamic limit, $n$ being the number of Hilbert space factors. On the other hand, for vacuum states of local theories $\Omega_{\textbf{A}}^{\epsilon}(\rho_{\textrm{vacuum}})$ is expected to be of $\mathcal{O}(1)$. Therefore, we conclude there is a hierarchical separation between both cases. This has a direct application in the case of expander graphs \cite{us3}. Even if the EE is globally extensive in the vacuum, $\Omega_{\textbf{A}}^{\epsilon}(\rho_{\textrm{vacuum}})$ is still expected to be of $\mathcal{O}(1)$.

In quantum mechanics, a theory of non-equilibrium thermal processes can be based on the assertion that Hamiltonian evolution drives an initial state to the sea of random/thermal/scrambled states. If the initial state is factorizable, we conclude that unitary evolution causes an irreversible growth of $\Omega_{\textbf{A}}^{\epsilon}(\rho)$ from $0$ to $\mathcal{O}(n)$. The non-equilibrium process codifies initially localized information in bigger and bigger subsystems until the maximum possible subsystem sizes of $\mathcal{O}(n)$. From this perspective, the information is not lost and might be found by looking in bigger subsystems, even though low point observables lose information and show effective thermality.

Besides, the functional form of the growth is intimately related with the structure of interactions of the theory, and the stabilization of $\Omega_{\textbf{A}}^{\epsilon}(\rho)$ gives a precise notion of the scrambling time \cite{hayden, susskind}.

We checked these claims in a non-integrable spin chain. The numerical simulations confirm  the theoretical expectations, including propagation bounds. We also compared the time average of the MI in the non-equilibrium processes with the ensemble average found analytically in Sec.~(\ref{sec:ex}), see Fig.  \ref{fig:therm}. The three cases are qualitatively the same, proving \emph{ergodicity} of MI in these systems.  The quantitative discrepancy was argued to be due to the different ensembles associated to each case.

As a final remark, notice that the Codification Volume seems to grow linearly for the case we studied, due to ballistic propagation of information. The process is expected to saturate after a time of $\mathcal{O}(n)$, providing the scrambling time for this type of systems. This ballistic propagation of information is then expected for tree like spin systems too. Concerning information transmission, trees are basically half lines/chains with a drift force in the direction of propagation \cite{us3,brownhypo}. However, for tree like structures this would imply an exponential growth of $\Omega_{ \textrm{1} }^{\epsilon}(\rho)$ as a function of time, due to the exponential growth of sites as we move radially in a tree like structure. Should this claim be correct, then quantum systems on trees or expander graphs would furnish specific examples of fast scramblers, providing the stationary value to  $\Omega_{ \textrm{1}}^{\epsilon}(\rho)$ of  $\mathcal{O}(n)$ in a time of $\mathcal{O}(\log n)$. Address this question will be a possible perspective for the future.

\section*{Acknowledgments}

We wish to thank J. L. F. Barb\'on, D. Campbell, P. Hayden, A. Queiroz and P. Sodano  for useful discussions. We acknowledge partial support from MCTI and UFRN/MEC (Brazil).

\addcontentsline{toc}{section}{Bibliography}

\bibliography{bibuscodif} 

\begin{thebibliography}{38}
\expandafter\ifx\csname natexlab\endcsname\relax\def\natexlab#1{#1}\fi
\expandafter\ifx\csname bibnamefont\endcsname\relax
  \def\bibnamefont#1{#1}\fi
\expandafter\ifx\csname bibfnamefont\endcsname\relax
  \def\bibfnamefont#1{#1}\fi
\expandafter\ifx\csname citenamefont\endcsname\relax
  \def\citenamefont#1{#1}\fi
\expandafter\ifx\csname url\endcsname\relax
  \def\url#1{\texttt{#1}}\fi
\expandafter\ifx\csname urlprefix\endcsname\relax\def\urlprefix{URL }\fi
\providecommand{\bibinfo}[2]{#2}
\providecommand{\eprint}[2][]{\url{#2}}

\bibitem[{\citenamefont{Deutsch}(1991)}]{Deutsch1991}
\bibinfo{author}{\bibfnamefont{J.~M.} \bibnamefont{Deutsch}},
  \bibinfo{journal}{Phys. Rev. A} \textbf{\bibinfo{volume}{43}},
  \bibinfo{pages}{2046} (\bibinfo{year}{1991}).

\bibitem[{\citenamefont{Page}(1993)}]{page}
\bibinfo{author}{\bibfnamefont{D.~N.} \bibnamefont{Page}},
  \bibinfo{journal}{Phys.Rev.Lett.} \textbf{\bibinfo{volume}{71}},
  \bibinfo{pages}{1291} (\bibinfo{year}{1993}).

\bibitem[{\citenamefont{Sen}(1996)}]{provepage}
\bibinfo{author}{\bibfnamefont{S.}~\bibnamefont{Sen}},
  \bibinfo{journal}{Phys.Rev.Lett} \textbf{\bibinfo{volume}{77}},
  \bibinfo{pages}{1} (\bibinfo{year}{1996}).

\bibitem[{\citenamefont{Srednicki}(1994)}]{srednicki1994}
\bibinfo{author}{\bibfnamefont{M.}~\bibnamefont{Srednicki}},
  \bibinfo{journal}{Phys. Rev. E} \textbf{\bibinfo{volume}{50}},
  \bibinfo{pages}{888} (\bibinfo{year}{1994}).

\bibitem[{\citenamefont{Hayden and Preskill}(2007)}]{hayden}
\bibinfo{author}{\bibfnamefont{P.}~\bibnamefont{Hayden}} \bibnamefont{and}
  \bibinfo{author}{\bibfnamefont{J.}~\bibnamefont{Preskill}},
  \bibinfo{journal}{JHEP} \textbf{\bibinfo{volume}{0709}}, \bibinfo{pages}{120}
  (\bibinfo{year}{2007}).

\bibitem[{\citenamefont{Sekino and Susskind}(2008)}]{susskind}
\bibinfo{author}{\bibfnamefont{Y.}~\bibnamefont{Sekino}} \bibnamefont{and}
  \bibinfo{author}{\bibfnamefont{L.}~\bibnamefont{Susskind}},
  \bibinfo{journal}{JHEP} \textbf{\bibinfo{volume}{0810}}, \bibinfo{pages}{065}
  (\bibinfo{year}{2008}).

\bibitem[{\citenamefont{Lashkari et~al.}(2013)\citenamefont{Lashkari, Stanford,
  Hastings, Osborne, and Hayden}}]{lashkari2013}
\bibinfo{author}{\bibfnamefont{N.}~\bibnamefont{Lashkari}},
  \bibinfo{author}{\bibfnamefont{D.}~\bibnamefont{Stanford}},
  \bibinfo{author}{\bibfnamefont{M.}~\bibnamefont{Hastings}},
  \bibinfo{author}{\bibfnamefont{T.}~\bibnamefont{Osborne}}, \bibnamefont{and}
  \bibinfo{author}{\bibfnamefont{P.}~\bibnamefont{Hayden}},
  \bibinfo{journal}{JHEP} \textbf{\bibinfo{volume}{1304}}, \bibinfo{pages}{022}
  (\bibinfo{year}{2013}).

\bibitem[{\citenamefont{Barbon and Magan}(2012)}]{us3}
\bibinfo{author}{\bibfnamefont{J.}~\bibnamefont{Barbon}} \bibnamefont{and}
  \bibinfo{author}{\bibfnamefont{J.}~\bibnamefont{Magan}},
  \bibinfo{journal}{JHEP} \textbf{\bibinfo{volume}{1208}}, \bibinfo{pages}{016}
  (\bibinfo{year}{2012}).

\bibitem[{\citenamefont{Shenker and Stanford}(2014)}]{stanford2014}
\bibinfo{author}{\bibfnamefont{S.}~\bibnamefont{Shenker}} \bibnamefont{and}
  \bibinfo{author}{\bibfnamefont{D.}~\bibnamefont{Stanford}},
  \bibinfo{journal}{Journal of High Energy Physics}
  \textbf{\bibinfo{volume}{2014}}, \bibinfo{pages}{67} (\bibinfo{year}{2014}).

\bibitem[{\citenamefont{Abajo-Arrastia
  et~al.}(2010)\citenamefont{Abajo-Arrastia, Apar\'icio, and
  L\'opez}}]{esperanza2010}
\bibinfo{author}{\bibfnamefont{J.}~\bibnamefont{Abajo-Arrastia}},
  \bibinfo{author}{\bibfnamefont{J.}~\bibnamefont{Apar\'icio}},
  \bibnamefont{and} \bibinfo{author}{\bibfnamefont{E.}~\bibnamefont{L\'opez}},
  \bibinfo{journal}{JHEP} \textbf{\bibinfo{volume}{11}}, \bibinfo{pages}{149}
  (\bibinfo{year}{2010}).

\bibitem[{\citenamefont{Balasubramanian
  et~al.}(2011)\citenamefont{Balasubramanian, Bernamonti, de~Boer, Copland,
  Craps, Keski-Vakkuri, M\"uller, Sch\"afer, Shigemori, and
  Staessens}}]{vijay2011}
\bibinfo{author}{\bibfnamefont{V.}~\bibnamefont{Balasubramanian}},
  \bibinfo{author}{\bibfnamefont{A.}~\bibnamefont{Bernamonti}},
  \bibinfo{author}{\bibfnamefont{J.}~\bibnamefont{de~Boer}},
  \bibinfo{author}{\bibfnamefont{N.}~\bibnamefont{Copland}},
  \bibinfo{author}{\bibfnamefont{B.}~\bibnamefont{Craps}},
  \bibinfo{author}{\bibfnamefont{E.}~\bibnamefont{Keski-Vakkuri}},
  \bibinfo{author}{\bibfnamefont{B.}~\bibnamefont{M\"uller}},
  \bibinfo{author}{\bibfnamefont{A.}~\bibnamefont{Sch\"afer}},
  \bibinfo{author}{\bibfnamefont{M.}~\bibnamefont{Shigemori}},
  \bibnamefont{and}
  \bibinfo{author}{\bibfnamefont{W.}~\bibnamefont{Staessens}},
  \bibinfo{journal}{Phys. Rev. D} \textbf{\bibinfo{volume}{84}},
  \bibinfo{pages}{026010} (\bibinfo{year}{2011}).

\bibitem[{\citenamefont{Tasaki}(1998)}]{Tasaki1998}
\bibinfo{author}{\bibfnamefont{H.}~\bibnamefont{Tasaki}},
  \bibinfo{journal}{Phys. Rev. Lett.} \textbf{\bibinfo{volume}{80}},
  \bibinfo{pages}{1373} (\bibinfo{year}{1998}).

\bibitem[{\citenamefont{Linden et~al.}(2009)\citenamefont{Linden, Popescu,
  Short, and Winter}}]{Linden2009}
\bibinfo{author}{\bibfnamefont{N.}~\bibnamefont{Linden}},
  \bibinfo{author}{\bibfnamefont{S.}~\bibnamefont{Popescu}},
  \bibinfo{author}{\bibfnamefont{A.~J.} \bibnamefont{Short}}, \bibnamefont{and}
  \bibinfo{author}{\bibfnamefont{A.}~\bibnamefont{Winter}},
  \bibinfo{journal}{Phys. Rev. E} \textbf{\bibinfo{volume}{79}},
  \bibinfo{pages}{061103} (\bibinfo{year}{2009}).

\bibitem[{\citenamefont{Riera et~al.}(2012)\citenamefont{Riera, Gogolin, and
  Eisert}}]{Riera2012}
\bibinfo{author}{\bibfnamefont{A.}~\bibnamefont{Riera}},
  \bibinfo{author}{\bibfnamefont{C.}~\bibnamefont{Gogolin}}, \bibnamefont{and}
  \bibinfo{author}{\bibfnamefont{J.}~\bibnamefont{Eisert}},
  \bibinfo{journal}{Phys. Rev. Lett.} \textbf{\bibinfo{volume}{108}},
  \bibinfo{pages}{080402} (\bibinfo{year}{2012}).

\bibitem[{\citenamefont{Calabrese and Cardy}(2006)}]{Calabrese2006}
\bibinfo{author}{\bibfnamefont{P.}~\bibnamefont{Calabrese}} \bibnamefont{and}
  \bibinfo{author}{\bibfnamefont{J.}~\bibnamefont{Cardy}},
  \bibinfo{journal}{Phys. Rev. Lett.} \textbf{\bibinfo{volume}{96}},
  \bibinfo{pages}{136801} (\bibinfo{year}{2006}).

\bibitem[{\citenamefont{Barthel and Schollw\"ock}(2008)}]{Barthel2008}
\bibinfo{author}{\bibfnamefont{T.}~\bibnamefont{Barthel}} \bibnamefont{and}
  \bibinfo{author}{\bibfnamefont{U.}~\bibnamefont{Schollw\"ock}},
  \bibinfo{journal}{Phys. Rev. Lett.} \textbf{\bibinfo{volume}{100}},
  \bibinfo{pages}{100601} (\bibinfo{year}{2008}).

\bibitem[{\citenamefont{Rossini et~al.}(2009)\citenamefont{Rossini, Silva,
  Mussardo, and Santoro}}]{mussardo2009}
\bibinfo{author}{\bibfnamefont{D.}~\bibnamefont{Rossini}},
  \bibinfo{author}{\bibfnamefont{A.}~\bibnamefont{Silva}},
  \bibinfo{author}{\bibfnamefont{G.}~\bibnamefont{Mussardo}}, \bibnamefont{and}
  \bibinfo{author}{\bibfnamefont{G.~E.} \bibnamefont{Santoro}},
  \bibinfo{journal}{Phys. Rev. Lett.} \textbf{\bibinfo{volume}{102}},
  \bibinfo{pages}{127204} (\bibinfo{year}{2009}).

\bibitem[{\citenamefont{Santos et~al.}(2012)\citenamefont{Santos, Polkovnikov,
  and Rigol}}]{lea2012}
\bibinfo{author}{\bibfnamefont{L.~F.} \bibnamefont{Santos}},
  \bibinfo{author}{\bibfnamefont{A.}~\bibnamefont{Polkovnikov}},
  \bibnamefont{and} \bibinfo{author}{\bibfnamefont{M.}~\bibnamefont{Rigol}},
  \bibinfo{journal}{Phys. Rev. E} \textbf{\bibinfo{volume}{86}},
  \bibinfo{pages}{010102} (\bibinfo{year}{2012}).

\bibitem[{\citenamefont{De~Pasquale et~al.}(2010)\citenamefont{De~Pasquale,
  Facchi, Parisi, Pascazio, and Scardicchio}}]{scardicchio2010}
\bibinfo{author}{\bibfnamefont{A.}~\bibnamefont{De~Pasquale}},
  \bibinfo{author}{\bibfnamefont{P.}~\bibnamefont{Facchi}},
  \bibinfo{author}{\bibfnamefont{G.}~\bibnamefont{Parisi}},
  \bibinfo{author}{\bibfnamefont{S.}~\bibnamefont{Pascazio}}, \bibnamefont{and}
  \bibinfo{author}{\bibfnamefont{A.}~\bibnamefont{Scardicchio}},
  \bibinfo{journal}{Phys. Rev. A} \textbf{\bibinfo{volume}{81}},
  \bibinfo{pages}{052324} (\bibinfo{year}{2010}).

\bibitem[{\citenamefont{Brand\~ao et~al.}(2012)\citenamefont{Brand\~ao,
  Cwiklinski, Horodecki, Horodecki, Korbicz, and Mozrzymas}}]{brandao2012}
\bibinfo{author}{\bibfnamefont{F.~G. S.~L.} \bibnamefont{Brand\~ao}},
  \bibinfo{author}{\bibfnamefont{P.}~\bibnamefont{Cwiklinski}},
  \bibinfo{author}{\bibfnamefont{M.}~\bibnamefont{Horodecki}},
  \bibinfo{author}{\bibfnamefont{P.}~\bibnamefont{Horodecki}},
  \bibinfo{author}{\bibfnamefont{J.~K.} \bibnamefont{Korbicz}},
  \bibnamefont{and}
  \bibinfo{author}{\bibfnamefont{M.}~\bibnamefont{Mozrzymas}},
  \bibinfo{journal}{Phys. Rev. E} \textbf{\bibinfo{volume}{86}},
  \bibinfo{pages}{031101} (\bibinfo{year}{2012}).

\bibitem[{\citenamefont{Huse and Oganesyan}(2013)}]{vadim}
\bibinfo{author}{\bibfnamefont{D.~A.} \bibnamefont{Huse}} \bibnamefont{and}
  \bibinfo{author}{\bibfnamefont{V.}~\bibnamefont{Oganesyan}}
  (\bibinfo{year}{2013}), \eprint{arXiv:1404.5216}.

\bibitem[{\citenamefont{Nanduri et~al.}(2014)\citenamefont{Nanduri, Kim, and
  Huse}}]{huse}
\bibinfo{author}{\bibfnamefont{A.}~\bibnamefont{Nanduri}},
  \bibinfo{author}{\bibfnamefont{H.}~\bibnamefont{Kim}}, \bibnamefont{and}
  \bibinfo{author}{\bibfnamefont{D.~A.} \bibnamefont{Huse}}
  (\bibinfo{year}{2014}), \eprint{arXiv:1404.5216}.

\bibitem[{\citenamefont{Pal and Huse}(2010)}]{huse2}
\bibinfo{author}{\bibfnamefont{A.}~\bibnamefont{Pal}} \bibnamefont{and}
  \bibinfo{author}{\bibfnamefont{D.~A.} \bibnamefont{Huse}},
  \bibinfo{journal}{Phys. Rev. B} \textbf{\bibinfo{volume}{82}},
  \bibinfo{pages}{174411} (\bibinfo{year}{2010}).

\bibitem[{\citenamefont{Oganesyan and Huse}(2007)}]{vadim3}
\bibinfo{author}{\bibfnamefont{V.}~\bibnamefont{Oganesyan}} \bibnamefont{and}
  \bibinfo{author}{\bibfnamefont{D.~A.} \bibnamefont{Huse}},
  \bibinfo{journal}{Phys. Rev. B} \textbf{\bibinfo{volume}{75}},
  \bibinfo{pages}{155111} (\bibinfo{year}{2007}).

\bibitem[{\citenamefont{Gogolin et~al.}(2011)\citenamefont{Gogolin, M\"uller,
  and Eisert}}]{Gogolin2011}
\bibinfo{author}{\bibfnamefont{C.}~\bibnamefont{Gogolin}},
  \bibinfo{author}{\bibfnamefont{M.~P.} \bibnamefont{M\"uller}},
  \bibnamefont{and} \bibinfo{author}{\bibfnamefont{J.}~\bibnamefont{Eisert}},
  \bibinfo{journal}{Phys. Rev. Lett.} \textbf{\bibinfo{volume}{106}},
  \bibinfo{pages}{040401} (\bibinfo{year}{2011}).

\bibitem[{\citenamefont{Cassidy et~al.}(2011)\citenamefont{Cassidy, Clark, and
  Rigol}}]{Cassidy2011}
\bibinfo{author}{\bibfnamefont{A.~C.} \bibnamefont{Cassidy}},
  \bibinfo{author}{\bibfnamefont{C.~W.} \bibnamefont{Clark}}, \bibnamefont{and}
  \bibinfo{author}{\bibfnamefont{M.}~\bibnamefont{Rigol}},
  \bibinfo{journal}{Phys. Rev. Lett.} \textbf{\bibinfo{volume}{106}},
  \bibinfo{pages}{140405} (\bibinfo{year}{2011}).

\bibitem[{\citenamefont{Hoory et~al.}(2006)\citenamefont{Hoory, Linial, and
  Wigderson}}]{avi1}
\bibinfo{author}{\bibfnamefont{S.}~\bibnamefont{Hoory}},
  \bibinfo{author}{\bibfnamefont{N.}~\bibnamefont{Linial}}, \bibnamefont{and}
  \bibinfo{author}{\bibfnamefont{A.}~\bibnamefont{Wigderson}},
  \bibinfo{journal}{Bull. Amer. Math. Soc. (N.S} \textbf{\bibinfo{volume}{43}},
  \bibinfo{pages}{439} (\bibinfo{year}{2006}).

\bibitem[{\citenamefont{{Lubotzky}}(2011)}]{avi2}
\bibinfo{author}{\bibfnamefont{A.}~\bibnamefont{{Lubotzky}}},
  \bibinfo{journal}{ArXiv e-prints}  (\bibinfo{year}{2011}),
  \eprint{1105.2389}.

\bibitem[{\citenamefont{Barbon and Fuertes}(2008)}]{fuertes}
\bibinfo{author}{\bibfnamefont{J.}~\bibnamefont{Barbon}} \bibnamefont{and}
  \bibinfo{author}{\bibfnamefont{C.}~\bibnamefont{Fuertes}},
  \bibinfo{journal}{JHEP} \textbf{\bibinfo{volume}{0804}}, \bibinfo{pages}{096}
  (\bibinfo{year}{2008}).

\bibitem[{\citenamefont{Balachandran
  et~al.}(2013{\natexlab{a}})\citenamefont{Balachandran, Govindarajan,
  de~Queiroz, and Reyes-Lega}}]{amilcar1}
\bibinfo{author}{\bibfnamefont{A.~P.} \bibnamefont{Balachandran}},
  \bibinfo{author}{\bibfnamefont{T.~R.} \bibnamefont{Govindarajan}},
  \bibinfo{author}{\bibfnamefont{A.~R.} \bibnamefont{de~Queiroz}},
  \bibnamefont{and} \bibinfo{author}{\bibfnamefont{A.~F.}
  \bibnamefont{Reyes-Lega}}, \bibinfo{journal}{Phys. Rev. Lett.}
  \textbf{\bibinfo{volume}{110}}, \bibinfo{pages}{080503}
  (\bibinfo{year}{2013}{\natexlab{a}}).

\bibitem[{\citenamefont{Balachandran
  et~al.}(2013{\natexlab{b}})\citenamefont{Balachandran, Govindarajan,
  de~Queiroz, and Reyes-Lega}}]{amilcar2}
\bibinfo{author}{\bibfnamefont{A.~P.} \bibnamefont{Balachandran}},
  \bibinfo{author}{\bibfnamefont{T.~R.} \bibnamefont{Govindarajan}},
  \bibinfo{author}{\bibfnamefont{A.~R.} \bibnamefont{de~Queiroz}},
  \bibnamefont{and} \bibinfo{author}{\bibfnamefont{A.~F.}
  \bibnamefont{Reyes-Lega}}, \emph{\bibinfo{title}{Algebraic approach to
  entanglement and entropy}} (\bibinfo{year}{2013}{\natexlab{b}}),
  \eprint{1301.1300}.

\bibitem[{\citenamefont{{Audenaert} and {Eisert}}(2005)}]{eisert}
\bibinfo{author}{\bibfnamefont{K.~M.~R.} \bibnamefont{{Audenaert}}}
  \bibnamefont{and} \bibinfo{author}{\bibfnamefont{J.}~\bibnamefont{{Eisert}}},
  \bibinfo{journal}{J. Math. Phys.} \textbf{\bibinfo{volume}{46}},
  \bibinfo{pages}{102104} (\bibinfo{year}{2005}).

\bibitem[{\citenamefont{Wolf et~al.}(2008)\citenamefont{Wolf, Verstraete,
  Hastings, and Cirac}}]{cirac}
\bibinfo{author}{\bibfnamefont{M.~M.} \bibnamefont{Wolf}},
  \bibinfo{author}{\bibfnamefont{F.}~\bibnamefont{Verstraete}},
  \bibinfo{author}{\bibfnamefont{M.~B.} \bibnamefont{Hastings}},
  \bibnamefont{and} \bibinfo{author}{\bibfnamefont{J.~I.} \bibnamefont{Cirac}},
  \bibinfo{journal}{Phys. Rev. Lett.} \textbf{\bibinfo{volume}{100}},
  \bibinfo{pages}{070502} (\bibinfo{year}{2008}).

\bibitem[{\citenamefont{Groisman et~al.}(2005)\citenamefont{Groisman, Popescu,
  and Winter}}]{groisman2005}
\bibinfo{author}{\bibfnamefont{B.}~\bibnamefont{Groisman}},
  \bibinfo{author}{\bibfnamefont{S.}~\bibnamefont{Popescu}}, \bibnamefont{and}
  \bibinfo{author}{\bibfnamefont{A.}~\bibnamefont{Winter}},
  \bibinfo{journal}{Phys. Rev. A} \textbf{\bibinfo{volume}{72}},
  \bibinfo{pages}{032317} (\bibinfo{year}{2005}).

\bibitem[{\citenamefont{Molina-Vilaplana and Sodano}(2011)}]{sodano2011}
\bibinfo{author}{\bibfnamefont{J.}~\bibnamefont{Molina-Vilaplana}}
  \bibnamefont{and} \bibinfo{author}{\bibfnamefont{P.}~\bibnamefont{Sodano}},
  \bibinfo{journal}{Journal of High Energy Physics}
  \textbf{\bibinfo{volume}{2011}}, \bibinfo{pages}{1} (\bibinfo{year}{2011}).

\bibitem[{\citenamefont{Ba\~nuls et~al.}(2011)\citenamefont{Ba\~nuls, Cirac,
  and Hastings}}]{strongthermal}
\bibinfo{author}{\bibfnamefont{M.~C.} \bibnamefont{Ba\~nuls}},
  \bibinfo{author}{\bibfnamefont{J.~I.} \bibnamefont{Cirac}}, \bibnamefont{and}
  \bibinfo{author}{\bibfnamefont{M.~B.} \bibnamefont{Hastings}},
  \bibinfo{journal}{Phys. Rev. Lett.} \textbf{\bibinfo{volume}{106}},
  \bibinfo{pages}{050405} (\bibinfo{year}{2011}).

\bibitem[{\citenamefont{Lieb and Robinson}(1972)}]{lieb}
\bibinfo{author}{\bibfnamefont{E.}~\bibnamefont{Lieb}} \bibnamefont{and}
  \bibinfo{author}{\bibfnamefont{D.}~\bibnamefont{Robinson}},
  \bibinfo{journal}{Communications in Mathematical Physics}
  \textbf{\bibinfo{volume}{28}}, \bibinfo{pages}{251} (\bibinfo{year}{1972}),
  ISSN \bibinfo{issn}{1432-0916}.

\bibitem[{\citenamefont{Monthus and Texier}(1996)}]{brownhypo}
\bibinfo{author}{\bibfnamefont{C.}~\bibnamefont{Monthus}} \bibnamefont{and}
  \bibinfo{author}{\bibfnamefont{C.}~\bibnamefont{Texier}},
  \bibinfo{journal}{Journal of Physics A: Mathematical and General}
  \textbf{\bibinfo{volume}{29}}, \bibinfo{pages}{2399} (\bibinfo{year}{1996}).

\end{thebibliography}
\end{document}